\documentstyle[aps,epsbox]{revtex}
\tighten
\begin{document}
%\draft

%======================================%
%<<<<<<<<<<<< TITLE PAGE >>>>>>>>>>>>>>%
%======================================%

\preprint{hep-th/0101xxx}
\title{Static solutions in the $R^4$ brane world}
\author{Shinji Mukohyama}
\address{
Department of Physics and Astronomy, University of Victoria\\ 
Victoria, BC, Canada V8W 3P6
}

\maketitle

%======================================%
%<<<<<<<<<<<<< ABSTRACT >>>>>>>>>>>>>>>% 
%======================================%

\begin{abstract} 

A simple five-dimensional brane world model is proposed, motivated 
by M-theory compactified on a six-dimensional manifold of small
radius and an $S^1/Z_2$ of large radius. We include a leading-order
higher curvature correction to the tree-level bulk action since in
brane world scenarios the curvature scale in the bulk may be
comparable to the five-dimensional Planck scale and, thus, higher
curvature corrections may become important. As a tractable model of 
the bulk theory we consider pure gravity including a
$($Ricci-scalar$)^4$-correction to the Einstein-Hilbert action. In
this model theory, after a conformal transformation to the Einstein
frame, we numerically obtain static solutions, each of which consists
of a positive tension brane and a negative tension brane. For these 
solutions, we obtain two relations between the warp factor and the
brane tensions. The existence of these relations implies that,
contrary to the original Randall-Sundrum model, the so called radion
is no longer a zero mode. We conclude that the tension of our brane
should be negative and that fine-tuning of the tension of both branes
is necessary for a large warp factor to explain the large hierarchy 
between the Planck scale and the electroweak scale. 

\end{abstract}


%======================================%
% Introduction
%======================================%

\section{Introduction}

The idea that our world may be a brane embedded in a higher dimensional 
spacetime has been attracting a great deal of physical interest. This
idea is often called the brane world scenario and, as suggested by
Randall and Sundrum~\cite{RS1,RS2}, may be able to explain the
large hierarchy between the Planck scale and the electroweak scale in
a natural way. Many aspects of the brane world scenario have been
investigated: for example, the effective four-dimensional Einstein's 
equation on a positive tension brane~\cite{SMS}, weak
gravity~\cite{gravity,gravity-stabilization}, black
holes~\cite{Black-hole}, inflating branes~\cite{Inflating-brane},
cosmologies~\cite{Cosmology1,Cosmology2,Friedmann,Perturbation,creation}, 
and so on.

In the original Randall-Sundrum brane world scenario, these authors
considered five-dimensional pure gravity described by the
Einstein-Hilbert action with a negative cosmological constant and two
$3$-branes with tension. Because of the negative cosmological constant
the five-dimensional bulk geometry is highly curved and the curvature 
scale is possibly of the order of the five-dimensional Planck scale,
while the induced geometries on the branes are flat, provided that
brane tensions are fine-tuned. Therefore, it is expected that quantum 
effects in the bulk may be important in the brane world scenario,
since quantum effects in curved spacetime usually become important
when the spacetime geometry is highly curved or when the causal
structure is non-trivial~\cite{BD}. In this connection, several
authors investigated quantum effects in the brane world 
scenario~\cite{GPT,Casimir,Mukohyama2000,HD,HKP}.

In particular, in ref.~\cite{Mukohyama2000} exact semiclassical
solutions representing a static brane world with two branes were
obtained by analyzing the semiclassical Einstein's equation in
five-dimensions with a negative cosmological constant and conformally 
invariant bulk matter fields. There, the following two types of
solutions were found. Type-(a): solution with a positive tension brane
and a negative tension brane. Type-(b): solution with two positive
tension branes. For each type of semiclassical solution, two relations 
between the warp factor and brane tensions were found: one giving the
warp factor as a function of the brane tensions and another giving a
relation between the brane tensions.

Although it is interesting that we could obtain analytic solutions
in the model of ref.~\cite{Mukohyama2000}, it seems that this model is 
not realistic enough. As far as the author knows, there is no
realization of conformally invariant bulk matter fields starting from 
M-theory or superstring theory. Nonetheless, it is expected that the
solutions in ref.~\cite{Mukohyama2000} may actually capture some
important features of quantum effects in the brane world
scenario. Hence, it is worth while to extend the analysis of
ref.~\cite{Mukohyama2000} to more realistic brane world models which
also take bulk quantum effects into account. For this purpose, one
would like to consider higher curvature corrections to the tree-level
bulk action. As discussed in the next section, $R^4$ corrections are
realistic from the point view of M-theory.

In this paper a simple five-dimensional brane world model is
proposed, motivated by M-theory compactified on a
six-dimensional manifold of small radius and an $S^1/Z_2$ of large
radius. We include the leading-order higher curvature correction to
the tree-level bulk action. As a tractable model of the bulk theory we 
consider pure gravity including a $($Ricci-scalar$)^4$-correction to
the Einstein-Hilbert action. In this model theory, after a conformal
transformation to the Einstein frame, we numerically obtain static
solutions, each of which consists of a positive tension brane and a
negative tension brane. For these solutions, we obtain two relations
between the warp factor and brane tensions. Those solutions and
relations are a close analogue of the type-(a) solutions and 
relations obtained in the model of ref.~\cite{Mukohyama2000}. On the
other hand, in the present model it will be shown that there is no
analogue of the type-(b) solutions. This fact might be considered to
be consistent with the suggestion of refs.~\cite{GPT,HKP} that the
type-(b) solutions are unstable.

This paper is organized as follows. In Sec.~\ref{sec:model} we
describe a simple brane world model which take bulk quantum effects
into account. In Sec.~\ref{sec:solution} we numerically obtain static
solutions in the model, and two relations between the warp factor and 
brane tensions are derived. Sec.~\ref{sec:summary} is devoted to a
summary of this paper.

%======================================%
% Model description
%======================================%
\section{Model description}
	\label{sec:model}

In this section we propose a simple brane world model, motivated
by  M-theory compactified on a six-dimensional compact manifold
(eg. Calabi-Yau manifold) of small radius and an $S^1/Z_2$ of large 
radius~\cite{HW}. In this situation, effectively we may consider a 
five-dimensional theory compactified on the $S^1/Z_2$. In the
five-dimensional bulk action, we shall consider a correction by a
$R^4$ term to the Einstein-Hilbert term~\footnote{
We, off course, include the Einstein-Hilbert term since it appears in
the tree-level effective action in eleven dimensions~\cite{CJS}.} 
since several calculations of higher-order corrections to the
effective action suggest that in eleven-dimensions $R^2$ terms do not
appear but $R^4$ terms may 
appear~\cite{GG,effective-action}~\footnote{
The importance of $R^4$ terms in M-theory was originally pointed out
in ref.~\cite{GG} and many authors showed evidences of
it~\cite{effective-action}. Possible cosmological consequences were
discussed in ref.~\cite{EKOY}.}. 
It is expected that the $R^4$ corrections
play important roles in brane world scenarios since the curvature
scale in the bulk may be comparable to the five-dimensional Planck
scale and, thus, higher curvature corrections cannot be neglected.

On the other hand, as for the action on branes, higher curvature
corrections are expected to be less important than those in the bulk
and can be neglected since curvature scale induced on branes (at least
on our brane) should be small compared to the Planck scale in low
energy. Nonetheless, motivated by the four- and ten-dimensional
effective theory induced on the fixed points of
$S^1/Z_2$~\cite{HW,LOW}, we may include $R^2$ corrections to brane
actions. In the following, we will explicitly see that the $R^2$
corrections do not play any roles in our analysis of static
solutions.

Since general higher curvature terms are difficult to analyze, for
simplicity, we shall consider Ricci scalars only. Furthermore, we
assume that the compactification from eleven dimensions to five
dimensions is properly stabilized and, for simplicity again, we do not
consider the corresponding moduli as dynamical fields in five
dimensions. Namely, in our analysis, we shall consider the following
action. 
%============< EQUATION >==============%
%
\begin{equation}
 I = \int_M d^5x\sqrt{-g}\left[\frac{1}{2\kappa^2}
	R_5+a\kappa^2R_5^4-\Lambda\right]
	+ \int_{\Sigma} d^4y\sqrt{-q}(b R_4^2-\lambda)
	+ \int_{\bar{\Sigma}} d^4\bar{y}\sqrt{-\bar{q}}
	(\bar{b} \bar{R}_4^2-\bar{\lambda}), 
\end{equation}
%======================================%
where $\kappa$ and $\Lambda$ are the five-dimensional gravitational
constant and cosmological constant, $a$, $b$ and $\bar{b}$ are
dimensionless constants, and $\lambda$ and $\bar{\lambda}$ are brane
tensions. The fixed-point hypersurface, or the world volume of a
$3$-brane, $\Sigma$ (or $\bar{\Sigma}$) is represented by $x^M=Z^M(y)$
(or $x^M=\bar{Z}^M(\bar{y})$, respectively) and the induced metric
$q_{\mu\nu}$ (or $\bar{q}_{\mu\nu}$, respectively) is defined by 
%============< EQUATION >==============%
%
\begin{eqnarray}
 q_{\mu\nu}(y) & = & e^M_{\mu}(y)e^N_{\nu}(y)g_{MN}|_{x=Z(y)},
	\nonumber \\
 e^M_{\mu}(y) & = &\frac{\partial Z^M(y)}{\partial y^{\mu}}
\end{eqnarray}
%======================================%
(or $\bar{q}_{\mu\nu}(\bar{y})
=\bar{e}^M_{\mu}(\bar{y})\bar{e}^N_{\nu}(\bar{y})
g_{MN}|_{x=\bar{Z}(\bar{y})}$,
$\bar{e}^M_{\mu}(\bar{y})
=\partial\bar{Z}^M(\bar{y})/\partial\bar{y}^{\mu}$, respectively). 
The Ricci scalars $R_5$, $R_4$ and $\bar{R}_4$ are of $g_{MN}$,
$q_{\mu\nu}$ and $\bar{q}_{\mu\nu}$, respectively.

Following ref.~\cite{Maeda}, we perform the conformal
transformation 
%============< EQUATION >==============%
%
\begin{eqnarray}
 \hat{g}_{MN} & = & e^{(1/\sqrt{3})\kappa\psi}g_{MN},
	\nonumber \\
 \kappa\psi & = &
	\frac{2}{\sqrt{3}}\ln(1+8a\kappa^4R_5^3)
\end{eqnarray}
%======================================%
to obtain the following expression. 
%============< EQUATION >==============%
%
\begin{eqnarray}
 I & = & \int_M d^5x\sqrt{-\hat{g}}\left[
	\frac{\hat{R}_5}{2\kappa^2}
	-\frac{1}{2}\hat{g}^{MN}\partial_M\psi\partial_N\psi
	-U(\psi)\right] \nonumber\\
 & & +\int_{\Sigma} d^4y\sqrt{-\hat{q}}\left[
	b\left(\hat{R}_4+\sqrt{3}\kappa\hat{D}^2\psi
	-\frac{1}{2}\kappa^2
	\hat{q}^{\mu\nu}\partial_{\mu}\psi\partial_{\nu}\psi
	\right)^2
	-f(\psi)\right]	\nonumber\\
 & & +\int_{\bar{\Sigma}} d^4\bar{y}\sqrt{-\hat{\bar{q}}}\left[
	\bar{b}\left(\hat{\bar{R}}_4+\sqrt{3}\kappa\hat{\bar{D}}^2\psi 
	-\frac{1}{2}\kappa^2
	\hat{\bar{q}}^{\mu\nu}\partial_{\mu}\psi\partial_{\nu}\psi
	\right)^2
	-\bar{f}(\psi)\right],
	\label{eqn:equivalent-action}
\end{eqnarray}
%======================================%
where the conformally transformed induced metrics $\hat{q}_{\mu\nu}$
and $\hat{\bar{q}}_{\mu\nu}$ are defined by 
%============< EQUATION >==============%
%
\begin{eqnarray}
 \hat{q}_{\mu\nu}(y) & = & 
	e^{(1/\sqrt{3})\kappa\psi(Z(y))}q_{\mu\nu}(y)
	=  e^M_{\mu}(y)e^N_{\nu}(y)\hat{g}_{MN}|_{x=Z(y)},
	\nonumber\\
 \hat{\bar{q}}_{\mu\nu}(\bar{y}) & = & 
	e^{(1/\sqrt{3})\kappa\psi(\bar{Z}(\bar{y}))}
	\bar{q}_{\mu\nu}(\bar{y})
	=  \bar{e}^M_{\mu}(\bar{y})\bar{e}^N_{\nu}(\bar{y})
	\hat{g}_{MN}|_{x=\bar{Z}(\bar{y})}, 
\end{eqnarray}
%======================================%
$\hat{D}$ and $\hat{\bar{D}}$ are four-dimensional covariant
derivatives compatible with $\hat{q}_{\mu\nu}$ and
$\hat{\bar{q}}_{\mu\nu}$, respectively, and the Ricci scalars
$\hat{R}_5$, $\hat{R}_4$ and $\hat{\bar{R}}_4$ are of $\hat{g}_{MN}$,
$\hat{q}_{\mu\nu}$ and $\hat{\bar{q}}_{\mu\nu}$, respectively. The
potential $U(\psi)$ and the functions $f(\psi)$ and $\bar{f}(\psi)$
are given by 
%============< EQUATION >==============%
%
\begin{eqnarray}
 U(\psi) & = & e^{-(5\sqrt{3}/6)\kappa\psi}
	\left[\Lambda
	+ (3/16)\kappa^{-10/3}a^{-1/3}
	(e^{(\sqrt{3}/2)\kappa\psi}-1)^{4/3}
	\right], \nonumber\\
 f(\psi) & = & e^{-(2/\sqrt{3})\kappa\psi}\lambda, \nonumber\\
 \bar{f}(\psi) & = & e^{-(2/\sqrt{3})\kappa\psi}\bar{\lambda}. 
	\label{eqn:U-fs}
\end{eqnarray}
%======================================%
To obtain the expression (\ref{eqn:equivalent-action}) we have assumed
that $8\kappa^4aR_5^3>-1$. For negative $\Lambda$, the potential can
be rewritten as 
%============< EQUATION >==============%
%
\begin{equation}
 U(\psi) = |\Lambda|e^{-(5\sqrt{3}/6)\kappa\psi}
	\left[-1+ \alpha(e^{(\sqrt{3}/2)\kappa\psi}-1)^{4/3}
	\right], \nonumber\\
\end{equation}
%======================================%
where
%============< EQUATION >==============%
%
\begin{equation}
 \alpha = (3/16)\kappa^{-10/3}a^{-1/3}|\Lambda|^{-1}.
\end{equation}
%======================================%
Since the $\hat{g}$-dependent part of the action
(\ref{eqn:equivalent-action}) is of the Einstein-Hilbert form, the
conformal frame in which the metric is $\hat{g}_{MN}$ is called 
{\it Einstein frame}. On the other hand, we shall call another
conformal frame in which the metric is $g_{MN}$  
{\it the original frame}.

In this paper, we assume that $\Lambda<0$ and consider a static
configuration with the ansatz 
%============< EQUATION >==============%
%
\begin{eqnarray}
 \hat{g}_{MN}dx^Mdx^N & = & 
	e^{-2A(w)}\eta_{\mu\nu}dx^{\mu}dx^{\nu} + dw^2, 
	\nonumber\\
 \psi & = & \psi(w).
	\label{eqn:ansatz}
\end{eqnarray}
%======================================%
This ansatz represents a general configuration with the
four-dimensional Poincar\'{e} invariance. Off course, the set of all
configurations with the four-dimensional Poincar\'{e} invariance in
the Einstein frame is equivalent to that in the original frame. With
this ansatz the curvature-squared term in the brane action does not
contribute to the equation of motion at all. Einstein's equation and
the field equation of the field $\psi$ are given by 
%============< EQUATION >==============%
%
\begin{eqnarray}
 3\frac{d^2A}{dw^2}-\kappa^2\left(\frac{d\psi}{dw}\right)^2 
	& = & 0,\nonumber\\
 6\left(\frac{dA}{dw}\right)^2-\kappa^2\left[
	\frac{1}{2}\left(\frac{d\psi}{dw}\right)^2-U(\psi)\right] 
	& = & 0,\nonumber\\
 e^{4A}\frac{d}{dw}\left(e^{-4A}\frac{d\psi}{dw}\right)-U'(\psi) 
	& = & 0. 
	\label{eqn:E-eq}
\end{eqnarray}
%======================================%
Note that the third equation is dependent of the first two equations
unless $d\psi/dw=0$ (the Bianchi identity), while the first equation
can also result from the second and the last equations unless
$dA/dw=0$. When we compactify the $w$-direction by $S^1/Z_2$ so that
$w\sim w+2L$ and that $w\sim -w$, there appears the following matching 
condition for $\psi$. 
%============< EQUATION >==============%
%
\begin{eqnarray}
 2\lim_{w\to +0}\frac{d\psi}{dw} & = & 
	\left. f'(\psi)\right|_{w=0},	\nonumber\\
 2\lim_{w\to L-0}\frac{d\psi}{dw} & = & 
	-\left.\bar{f}'(\psi)\right|_{w=L}, 
	\label{eqn:matching-cond}
\end{eqnarray}
%======================================%
where we have assumed that $\Sigma$ and $\bar{\Sigma}$ are
world-volumes of branes at the two fixed points $w=0$ and $w=L$,
respectively. We suppose that the brane at $w=0$ is our world and
shall call it {\it our brane}. We shall call another brane at $w=L$
{\it the hidden brane}. As for the function $A(w)$ in the metric, we
have the junction condition 
%============< EQUATION >==============%
%
\begin{eqnarray}
 6\lim_{w\to +0}\frac{dA}{dw} & = & 
	\left.\kappa^2 f(\psi)\right|_{w=0},
	\nonumber\\
 6\lim_{w\to L-0}\frac{dA}{dw} & = & 
	-\left.\kappa^2\bar{f}(\psi)\right|_{w=L}.
	\label{eqn:junction-cond}
\end{eqnarray}
%======================================%
This is a special case of Israel's junction
condition~\cite{Israel}. In the Einstein frame the so called warp 
factor can be defined by $\phi_E=e^{A(0)}/e^{A(L)}$. Correspondingly,
the warp factor in the original frame is
%============< EQUATION >==============%
%
\begin{equation}
 \phi = \exp\left\{[A(0)-A(L)]
	+\frac{\kappa}{2\sqrt{3}}[\psi(0)-\psi(L)]\right\}.
	\label{eqn:warp-original}
\end{equation}
%======================================%

It is evident that $\psi\equiv \psi_0$ is not a solution because of
the matching condition (\ref{eqn:matching-cond}), where $\psi_0$ is an 
extremum of the potential $U(\psi)$. In particular, we can show that
%============< EQUATION >==============%
%
\begin{equation}
 \lim_{w\to +0}U(\psi) = \lim_{w\to L-0}U(\psi) = 0,
	\label{eqn:U=0}
\end{equation}
%======================================%
and thus $\psi$ cannot stay at $\psi_0$ unless $\Lambda$ is zero. 
Actually, provided that equations of motion (\ref{eqn:E-eq}) and the 
junction condition (\ref{eqn:junction-cond}) are satisfied, the
matching condition (\ref{eqn:matching-cond}) is equivalent to the
vanishing-potential condition (\ref{eqn:U=0}) combined with
%============< EQUATION >==============%
%
\begin{eqnarray}
 \lim_{w\to+0}\frac{dA}{dw}\cdot\frac{d\psi}{dw} & \leq & 0, 
	\nonumber\\
 \lim_{w\to L-0}\frac{dA}{dw}\cdot\frac{d\psi}{dw} & \leq & 0. 
	\label{eqn:dAdphi-cond}
\end{eqnarray}
%======================================%
In order to see the necessity of the condition
(\ref{eqn:dAdphi-cond}), notice that $f(\psi)f'(\psi)\leq 0$ and
$\bar{f}(\psi)\bar{f}'(\psi)\leq 0$.

%======================================%
% Numerical solution
%======================================%
\section{Numerical solution}
	\label{sec:solution}

\begin{figure}
 \begin{center}
  \epsfile{file=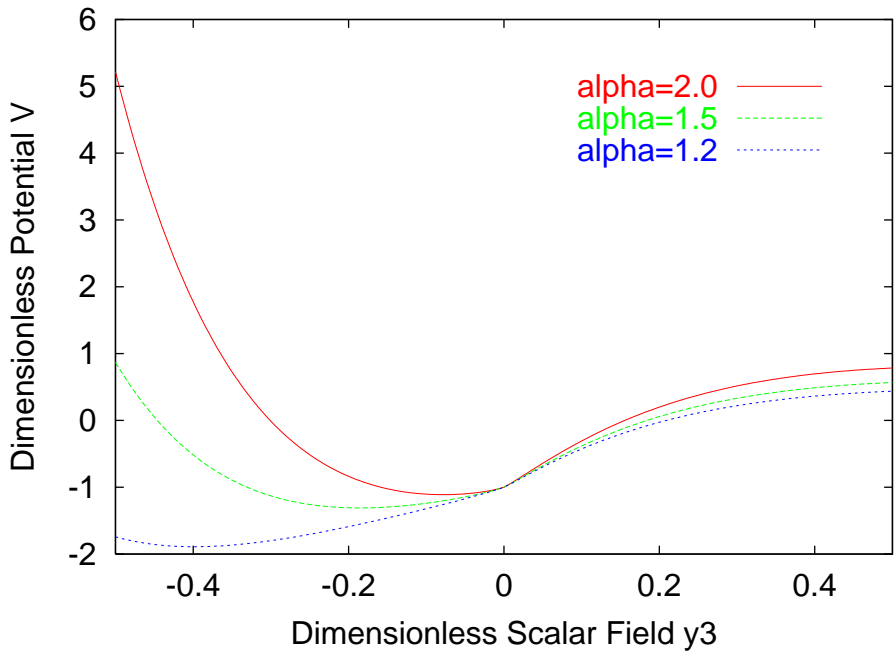,scale=1.0}
 \end{center}
\caption{
For $\alpha>1$ the dimensionless potential $V(y_3)$ has roots
$y_3=y_{\pm}$, where $y_{\pm}$ are given by (\ref{eqn:roots}), and a
global minimum. In this figure, $V(y_3)$ is shown for $\alpha=2.0$,
$1.5$, $1.2$. 
}
\label{fig:potential}
\end{figure}

For the purpose of numerical integration, it is convenient to rewrite
all equations in terms of dimensionless variables. Hence, we introduce 
the dimensionless independent variable $x$ defined by $x=w/L$ and
consider the region $0\leq x\leq 1$, where $L$ is the distance between
two branes. As for the dependent variables, we introduce the following 
three: 
%============< EQUATION >==============%
%
\begin{eqnarray}
 y_1(x) & \equiv & A, \nonumber\\
 y_2(x) & \equiv & L\frac{dA}{dw}, \nonumber\\
 y_3(x) & \equiv & \frac{\kappa}{2\sqrt{3}}\psi. 
\end{eqnarray}
%======================================%
Differential equations for these dimensionless independent variables
are given by 
%============< EQUATION >==============%
%
\begin{eqnarray}
 \dot{y}_1 & = & y_2, \nonumber\\
 \dot{y}_2 & = & 4[y_2^2+(L/l)^2V(y_3)], \nonumber\\
 (\dot{y}_3)^2 & = & y_2^2+(L/l)^2V(y_3),
	\label{eqn:diff-eq}
\end{eqnarray}
%======================================%
where dots denote differentiation with respect to $x$, the length
scale $l$ is defined by $l=\kappa^{-1}\sqrt{6/|\Lambda|}$, and 
%============< EQUATION >==============%
%
\begin{equation}
 V(y_3) = e^{-5y_3}\left[-1+\alpha(e^{3y_3}-1)^{4/3}\right]. 
\end{equation}
%======================================%
As already mentioned in the previous section we assume that
$\Lambda<0$. The set of these three differential equations is
equivalent to the equation of motion (\ref{eqn:E-eq}) as long as
$\dot{y}_2$ is not zero. As we shall argue later, there is no static
solution for $\alpha\leq 1$. Hence, for the present we shall
concentrate on the case $\alpha>1$ only. The potential is shown in 
Figure~\ref{fig:potential} for $\alpha=2.0$, $1.5$, and $1.2$. The
vanishing-potential condition (\ref{eqn:U=0}) is written as 
%============< EQUATION >==============%
%
\begin{equation}
 V(y_3(0)) = V(y_3(1)) = 0, \label{eqn:V=0}
\end{equation}
%======================================%
and should be complemented by 
%============< EQUATION >==============%
%
\begin{eqnarray}
 y_2(0)\dot{y}_3(0) & \leq & 0, \nonumber\\
 y_2(1)\dot{y}_3(1) & \leq & 0.	
	\label{eqn:complementary-cond}
\end{eqnarray}
%======================================%
It is easy to impose the boundary condition (\ref{eqn:V=0}) since the
roots of $V(y_3)$ are analytically obtained as $y_3=y_{\pm}$ for
$\alpha>1$, where
%============< EQUATION >==============%
%
\begin{equation}
 y_{\pm} = \frac{1}{3}\ln(1\pm\alpha^{-3/4}). 
	\label{eqn:roots}
\end{equation}
%======================================%
The complementary condition (\ref{eqn:complementary-cond}) should be
checked after a solution of the differential equation
(\ref{eqn:diff-eq}) with the boundary condition (\ref{eqn:V=0}) is
obtained. Thence, the junction condition (\ref{eqn:junction-cond})
determines the brane tensions as 
%============< EQUATION >==============%
%
\begin{eqnarray}
 \lambda/(6\kappa^{-2}l^{-1}) & = & 
	(l/L)y_2(0)e^{4y_3(0)}, \nonumber\\
 \bar{\lambda}/(6\kappa^{-2}l^{-1}) & = & 
	-(l/L)y_2(1)e^{4y_3(1)}. 
	\label{eqn:brane-tension}
\end{eqnarray}
%======================================%
Finally, the warp factor $\phi$ given by (\ref{eqn:warp-original}) is
written as
%============< EQUATION >==============%
%
\begin{equation}
 \phi = \exp\left[y_1(0)+y_3(0)-y_1(1)-y_3(1)\right]. 
	\label{eqn:warp-factor}
\end{equation}
%======================================%

%============< FIGURE >==============%
%	plot-potential.eps
%======================================%

Note that, without loss of generality, we can impose the additional
condition 
%============< EQUATION >==============%
%
\begin{equation}
 y_1(0) = 0,
	\label{eqn:y10=0}
\end{equation}
%======================================%
since none of the above equations is changed by the shift 
$y_1(x)\to y_1(x)-y_1(0)$. This additional condition combined with the 
vanishing-potential condition (\ref{eqn:V=0}), which can be rewritten
as 
%============< EQUATION >==============%
%
\begin{eqnarray}
 y_3(0) & = & y_{\pm}, \nonumber\\
 y_3(1) & = & y_{\pm}, \label{eqn:y3=root}
\end{eqnarray}
%======================================%
give enough number of boundary conditions for the set of three
differential equations (\ref{eqn:diff-eq}). Here, plus or minus signs
in two of (\ref{eqn:y3=root}) are independent. According to four
possible choices of the signs in (\ref{eqn:y3=root}), there are four
possible types of solutions. 
%============< EQUATION >==============%
%
\begin{eqnarray}
 (++)\mbox{-type} & : & y_3(0)=y_+,\ y_3(1)=y_+,\nonumber\\
 (+-)\mbox{-type} & : & y_3(0)=y_+,\ y_3(1)=y_-,\nonumber\\
 (-+)\mbox{-type} & : & y_3(0)=y_-,\ y_3(1)=y_+,\nonumber\\
 (--)\mbox{-type} & : & y_3(0)=y_-,\ y_3(1)=y_-. 
\end{eqnarray}
%======================================%

We shall solve the differential equation (\ref{eqn:diff-eq}) with the
boundary condition (\ref{eqn:y10=0}) and (\ref{eqn:y3=root}) by the so
called relaxation method~\cite{NR}. This method works very well if a
good initial guess is given. In the following we shall solve the
differential equation many times, each time with different values of
the parameters $\alpha$ and $L/l$. Hence, the previous solution can 
be used as a good initial guess in the next calculation with slightly
different parameters.

\begin{figure}
 \begin{center}
  \epsfile{file=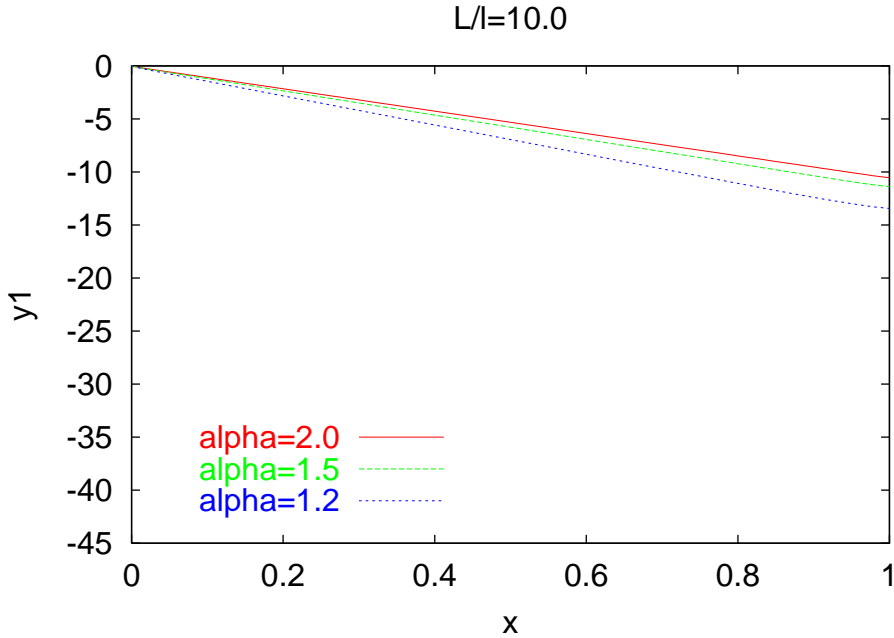,scale=1.0}
 \end{center}
\caption{
The numerical solution $y_1(x)$ for $L/l=10.0$ and $\alpha=2.0$,
$1.5$, $1.2$. 
}
	\label{fig:y1-10}
\end{figure}

\begin{figure}
 \begin{center}
  \epsfile{file=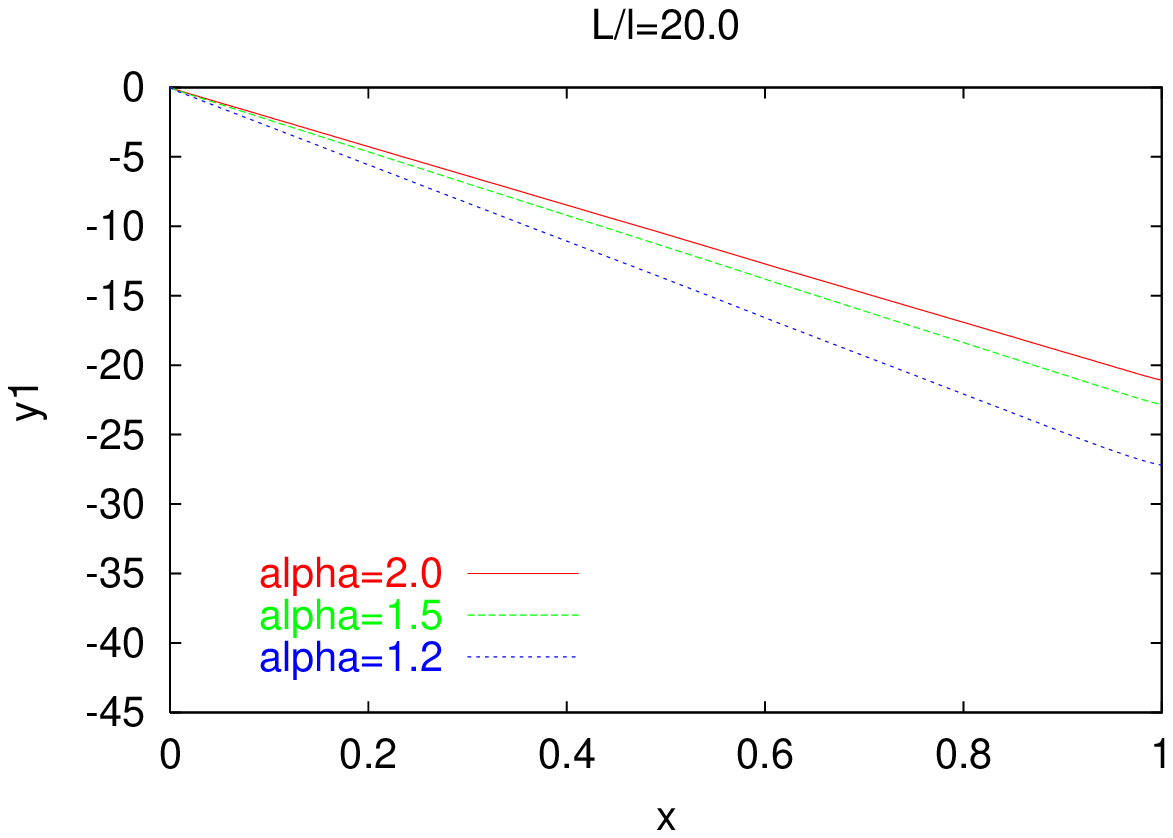,scale=1.0}
 \end{center}
\caption{
The numerical solution $y_1(x)$ for $L/l=20.0$ and $\alpha=2.0$,
$1.5$, $1.2$. 
}
	\label{fig:y1-20}
\end{figure}

\begin{figure}
 \begin{center}
  \epsfile{file=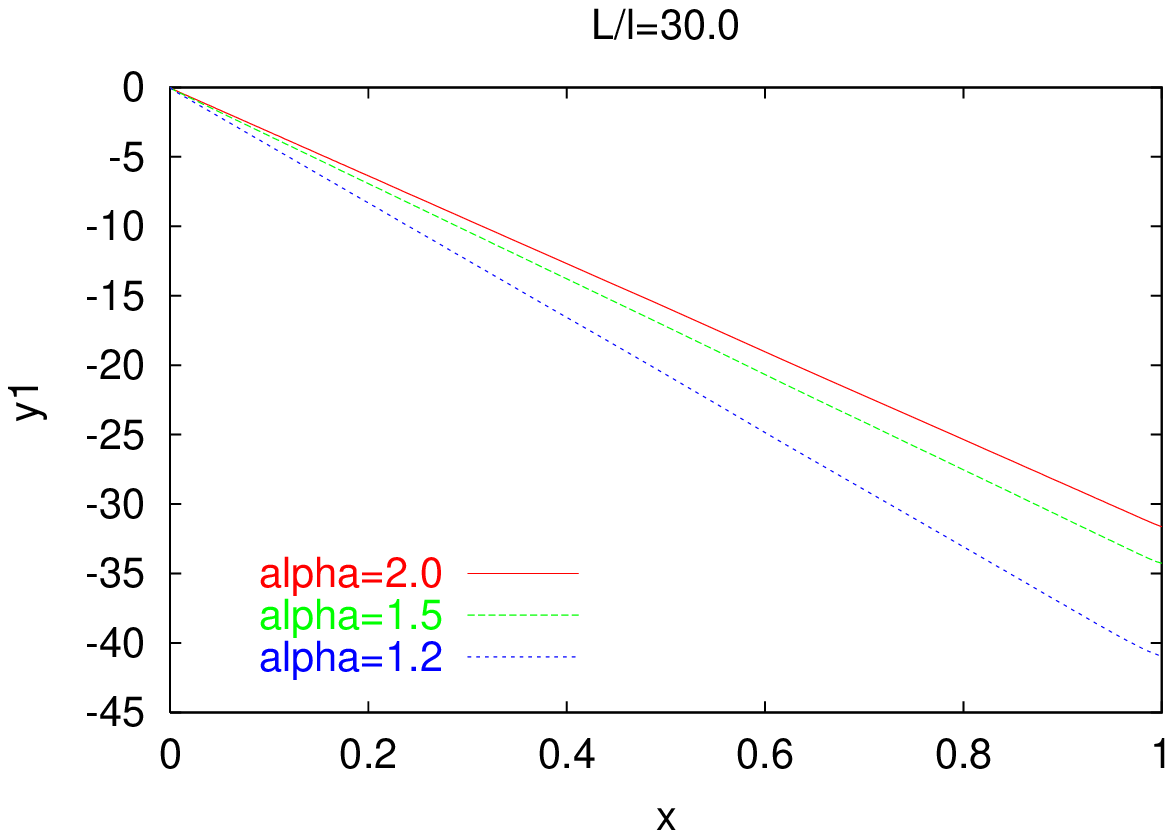,scale=1.0}
 \end{center}
\caption{
The numerical solution $y_1(x)$ for $L/l=30.0$ and $\alpha=2.0$,
$1.5$, $1.2$. 
}
	\label{fig:y1-30}
\end{figure}

\begin{figure}
 \begin{center}
  \epsfile{file=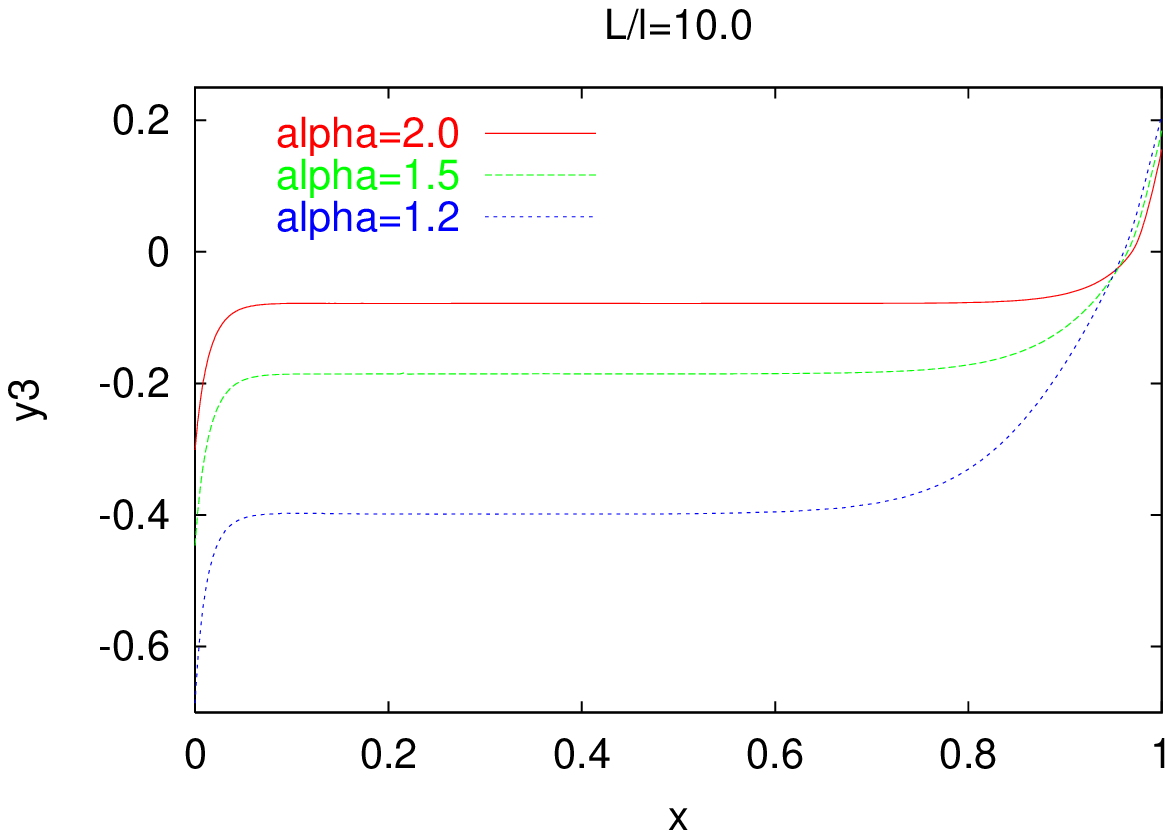,scale=1.0}
 \end{center}
\caption{
The numerical solution $y_3(x)$ for $L/l=10.0$ and $\alpha=2.0$,
$1.5$, $1.2$. 
}
	\label{fig:y3-10}
\end{figure}

\begin{figure}
 \begin{center}
  \epsfile{file=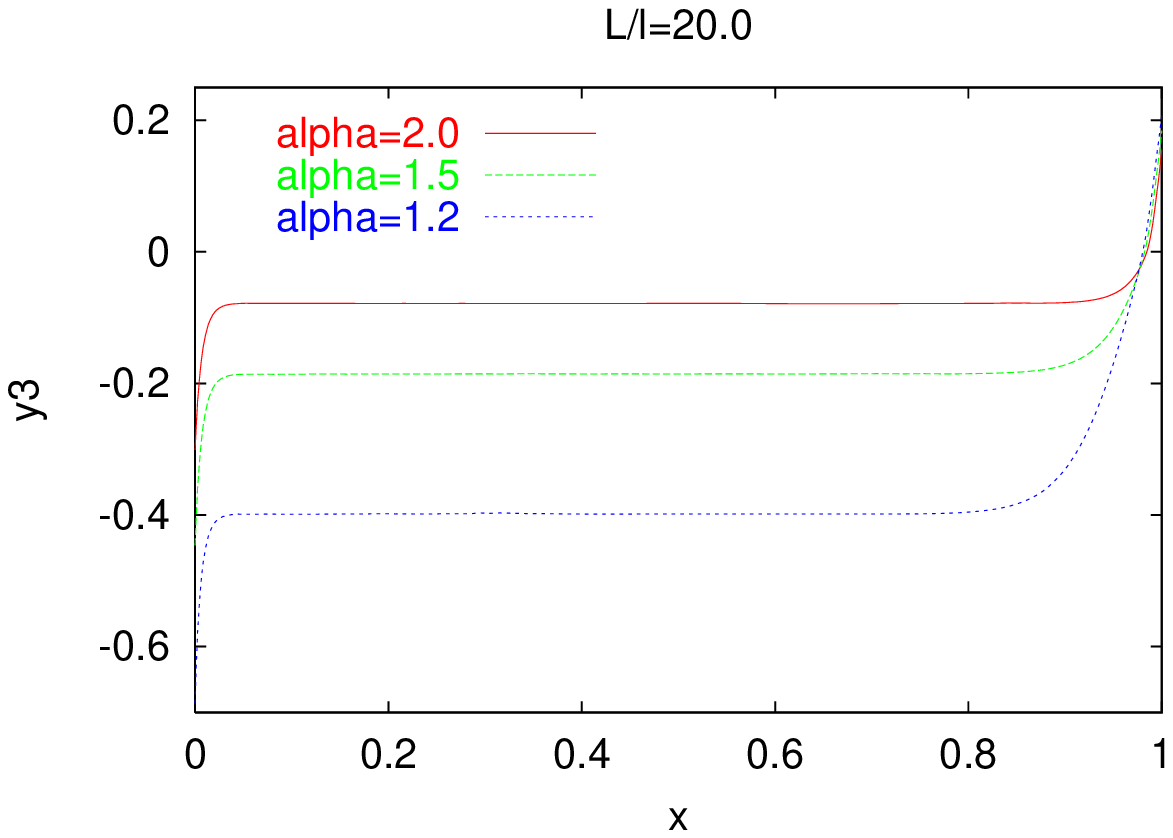,scale=1.0}
 \end{center}
\caption{
The numerical solution $y_3(x)$ for $L/l=20.0$ and $\alpha=2.0$,
$1.5$, $1.2$. 
}
	\label{fig:y3-20}
\end{figure}

\begin{figure}
 \begin{center}
  \epsfile{file=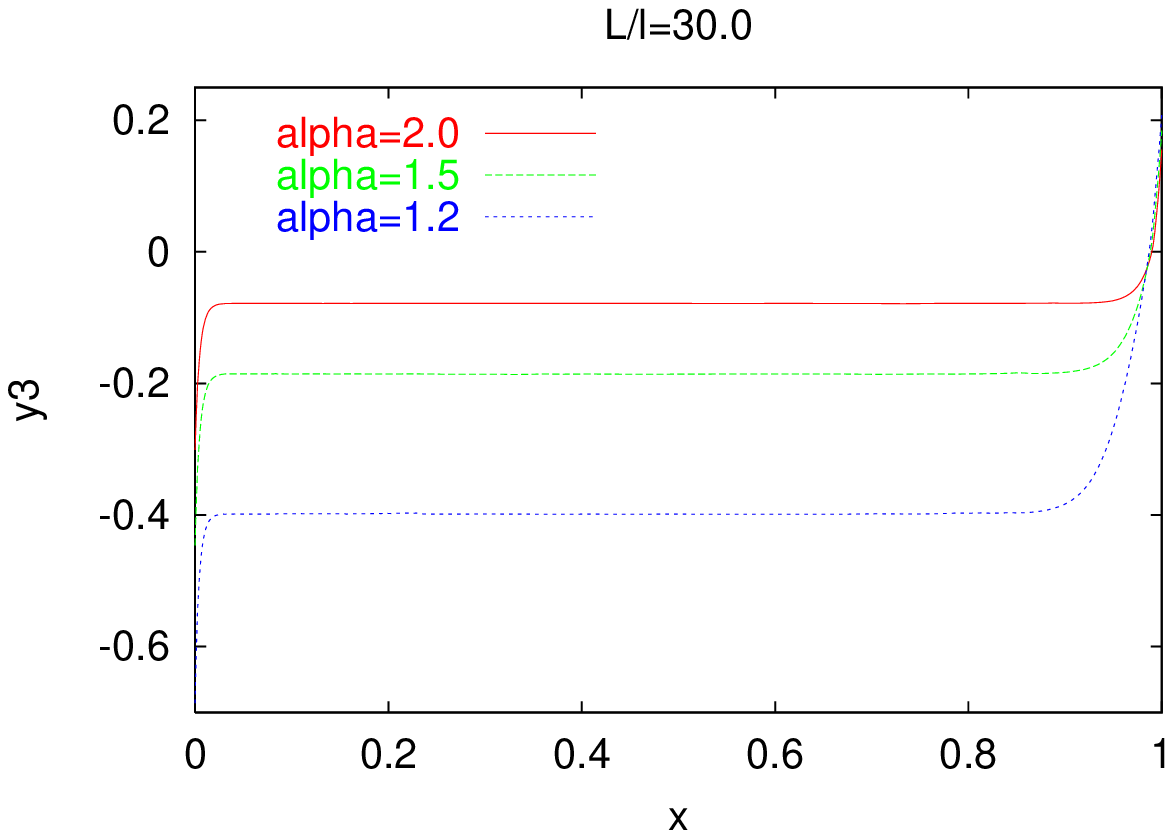,scale=1.0}
 \end{center}
\caption{
The numerical solution $y_3(x)$ for $L/l=30.0$ and $\alpha=2.0$,
$1.5$, $1.2$. 
}
	\label{fig:y3-30}
\end{figure}

Figures~\ref{fig:y1-10}, \ref{fig:y1-20}, \ref{fig:y1-30},
\ref{fig:y3-10}, \ref{fig:y3-20} and \ref{fig:y3-30} show several
$(-+)$-type solutions obtained by the relaxation method. Physical
parameters in these solutions are $\alpha=2.0$, $1.5$, $1.2$ and
$L/l=10.0$, $20.0$, $30.0$.

%============< FIGURE >==============%
%	plot-y1-10.eps
%======================================%

%============< FIGURE >==============%
%	plot-y1-20.eps
%======================================%

%============< FIGURE >==============%
%	plot-y1-30.eps
%======================================%

%============< FIGURE >==============%
%	plot-y3-10.eps
%======================================%

%============< FIGURE >==============%
%	plot-y3-20.eps
%======================================%

%============< FIGURE >==============%
%	plot-y3-30.eps
%======================================%

\begin{figure}
 \begin{center}
  \epsfile{file=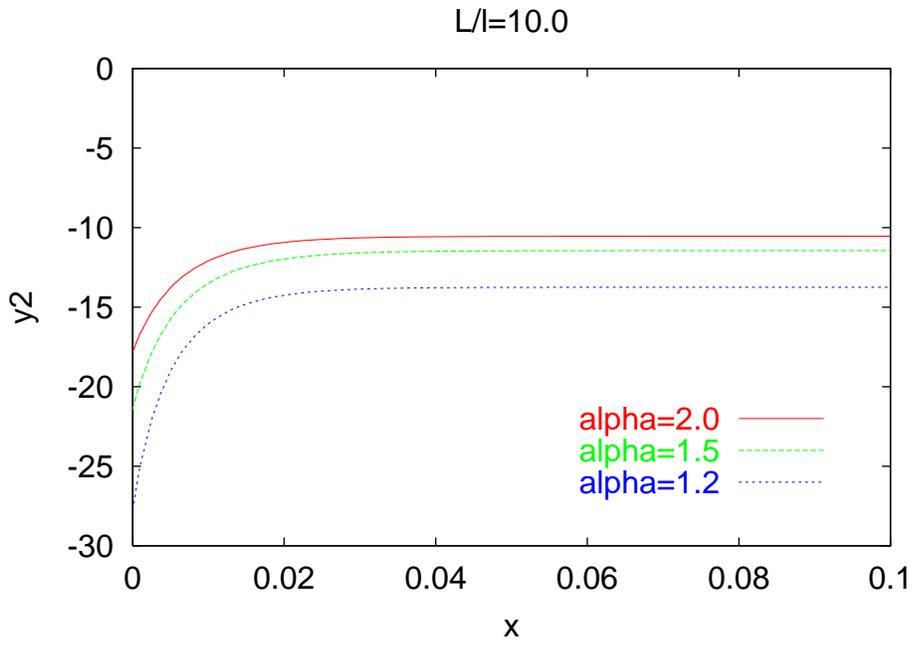,scale=1.0}
 \end{center}
\caption{
The numerical solution $y_2(x)\equiv\dot{y}_1(x)$ in the vicinity of
the boundary $x=0$ for $L/l=10.0$ and $\alpha=2.0$, $1.5$, $1.2$. 
}
	\label{fig:left-y2-10}
\end{figure}

\begin{figure}
 \begin{center}
  \epsfile{file=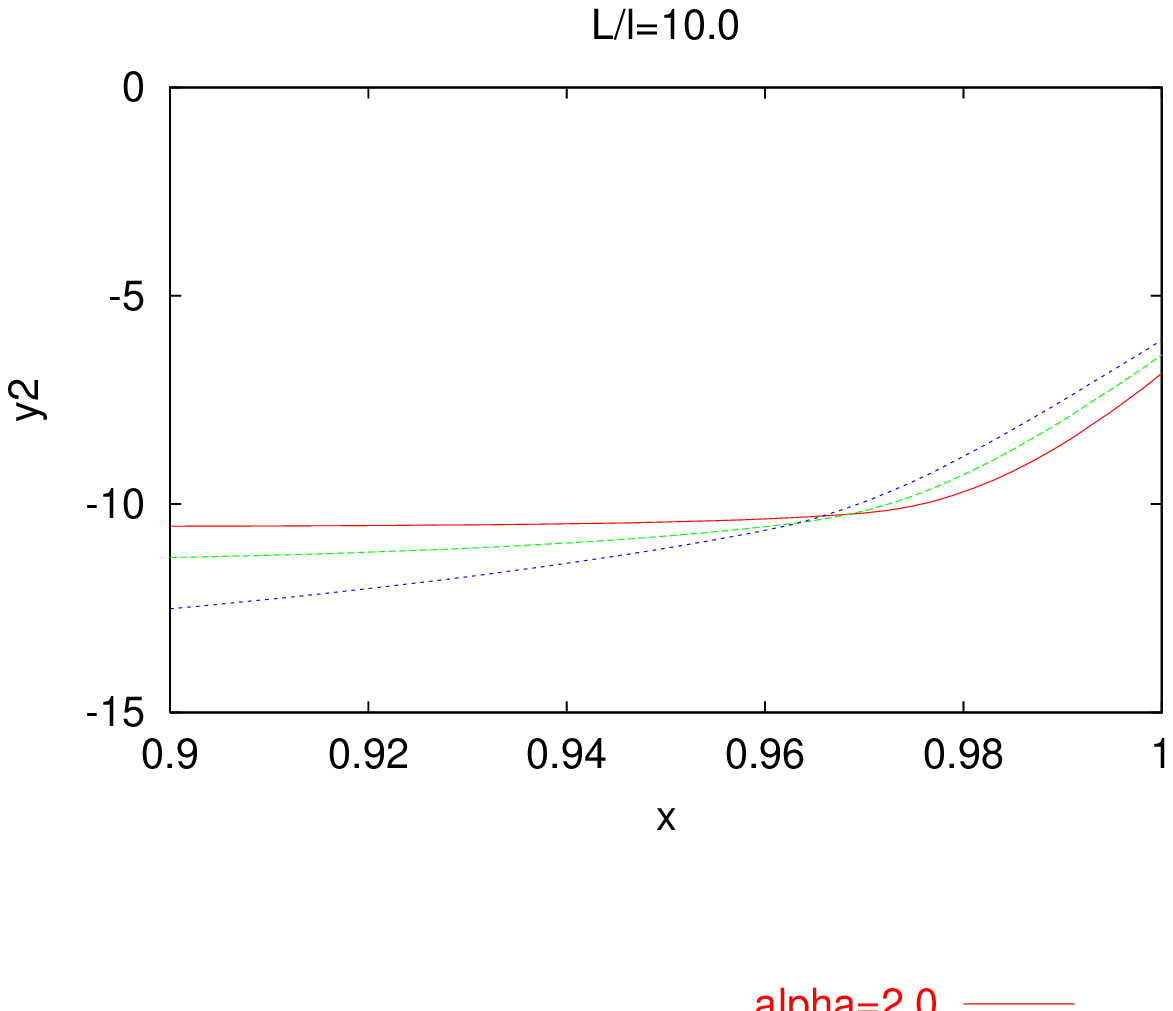,scale=1.0}
 \end{center}
\caption{
The numerical solution $y_2(x)\equiv\dot{y}_1(x)$ in the vicinity of
the boundary $x=1$ for $L/l=10.0$ and $\alpha=2.0$, $1.5$, $1.2$. 
}
	\label{fig:right-y2-10}
\end{figure}

\begin{figure}
 \begin{center}
  \epsfile{file=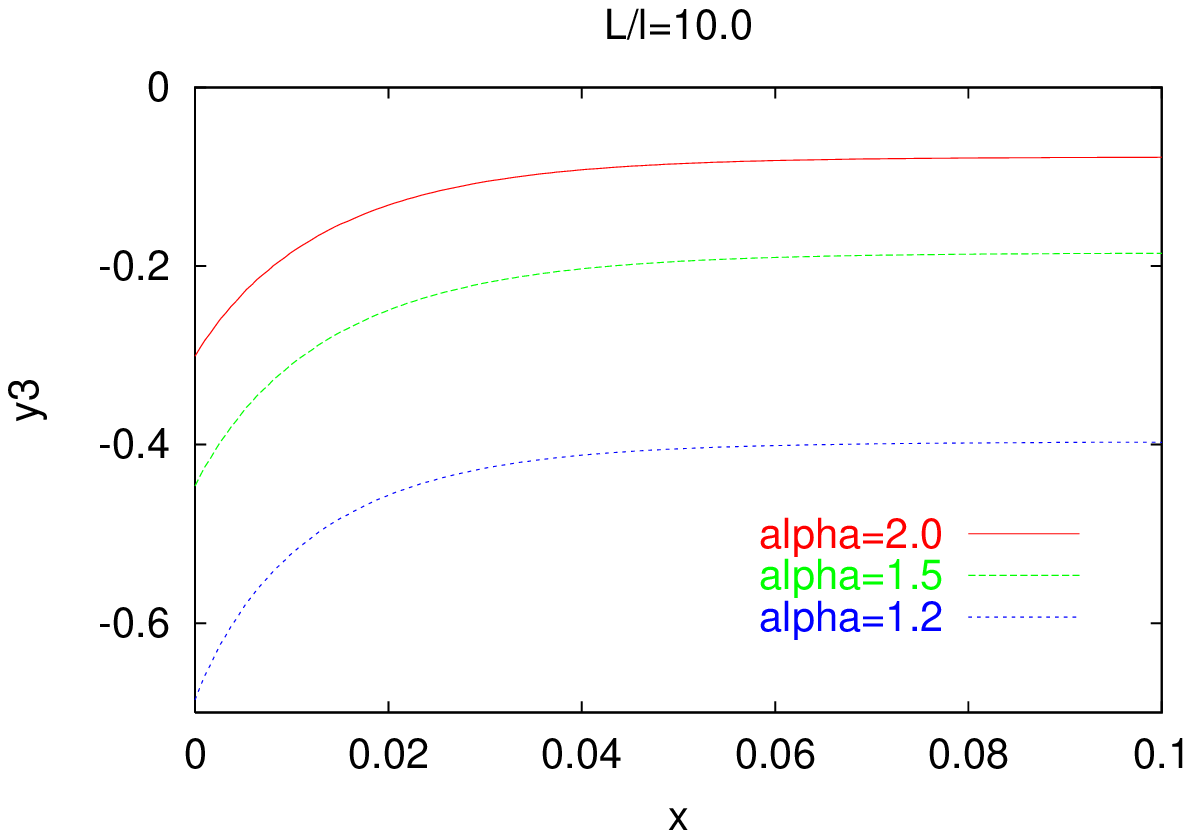,scale=1.0}
 \end{center}
\caption{
The numerical solution $y_3(x)$ in the vicinity of the boundary $x=0$
for $L/l=10.0$ and $\alpha=2.0$, $1.5$, $1.2$. 
}
	\label{fig:left-y3-10}
\end{figure}

\begin{figure}
 \begin{center}
  \epsfile{file=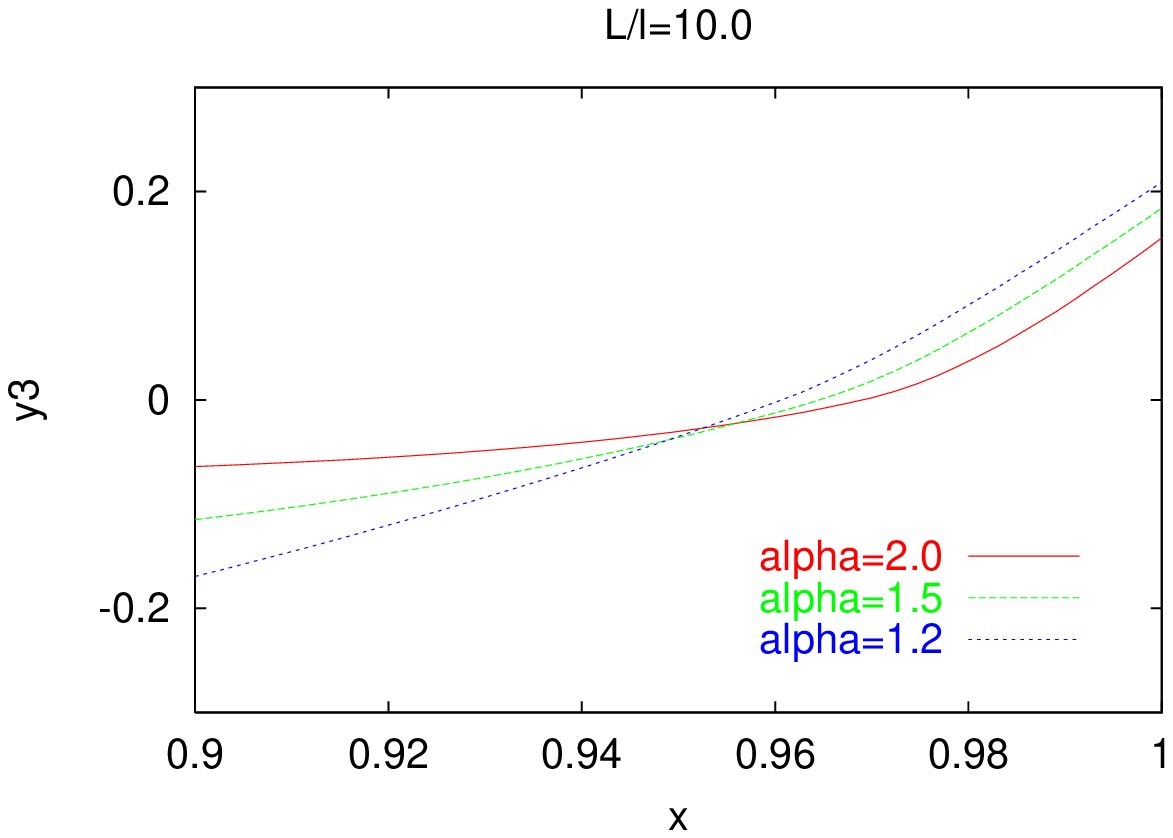,scale=1.0}
 \end{center}
\caption{
The numerical solution $y_3(x)$ in the vicinity of the boundary $x=1$
for $L/l=10.0$ and $\alpha=2.0$, $1.5$, $1.2$. 
}
	\label{fig:right-y3-10}
\end{figure}

From these figures, we can see that, except for vicinities of the
boundaries, $y_3$ stays near the minimum $y_{min}$ of the potential
$V(y_3)$ and $y_1$ is almost linear in $x$:~\footnote{
As we shall argue in the paragraph after the next, these approximate
expressions could be inferred without any numerical calculations for
large values of $L/l$.}
%============< EQUATION >==============%
%
\begin{eqnarray}
 y_1(x) & \simeq & -\frac{L}{l}\sqrt{|V(y_{min})|}\ x, \nonumber\\
 y_3(x) & \simeq & y_{mim}. \label{eqn:approx-y1-y3}
\end{eqnarray}
%======================================%
In other words, except for vicinities of the boundaries, the scalar
field $\psi$ stays near the minimum of the potential $U(\psi)$ and the
five-dimensional geometry is almost the AdS whose curvature is
determined by the minimal value of $U(\psi)$. However, because of the
boundary condition (\ref{eqn:y3=root}), near-boundary behaviors of
solutions are non-trivial. Actually, figures~\ref{fig:left-y2-10} and 
\ref{fig:right-y2-10} show that the five dimensional geometry near the 
boundaries deviates rather strongly from AdS. The deviation is larger
for a smaller value of $\alpha$, or a larger value of the coefficient
$a$ of the $R^4$-term, as easily expected. Note that $y_2=\dot{y}_1$
would be independent of $x$ if the geometry was AdS. Each of
figures~\ref{fig:left-y3-10} and \ref{fig:right-y3-10} shows how the
scalar field approaches one of roots of the potential. 
%============< FIGURE >==============%
%	left-closeplot-y2-10.eps
%======================================%

%============< FIGURE >==============%
%	right-closeplot-y2-10.eps
%======================================%

%============< FIGURE >==============%
%	left-closeplot-y3-10.eps
%======================================%

%============< FIGURE >==============%
%	right-closeplot-y3-10.eps
%======================================%

We now argue that there are no (static) solutions of ($++$)- and
($--$)-types. First, let us rewrite the differential equation as
%============< EQUATION >==============%
%
\begin{eqnarray}
 \dot{y}_1 & = & y_2, \nonumber\\
 \dot{y}_2 & = & 4y_4^2, \nonumber\\
 \dot{y}_3 & = & y_4, \nonumber\\
 \dot{y}_4 & = & 4y_2y_4+(L/l)^2V'(y_3)/2,
        \label{eqn:diff-eq-RK}
\end{eqnarray}
%======================================%
where
%============< EQUATION >==============%
%
\begin{equation}
 y_4(x) = \frac{\kappa}{2\sqrt{3}}L\frac{d\psi}{dw},
\end{equation}
%======================================%
and the corresponding boundary condition is 
%============< EQUATION >==============%
%
\begin{eqnarray}
 y_1(0) & = & 0, \nonumber\\
 y_2(0)+y_4(0) & = & 0, \nonumber\\
 y_3(0) & = & y_{\pm}, \nonumber\\
 y_2(1)+y_4(1) & = & 0. 
        \label{eqn:bc-RK}
\end{eqnarray}
%======================================%
It is easy to show from (\ref{eqn:diff-eq-RK}) and (\ref{eqn:bc-RK})
that 
%============< EQUATION >==============%
%
\begin{equation}
 \int_0^1 dxe^{-4y_3}V'(y_3) =0. 
	\label{eqn:deV=0}
\end{equation}
%======================================%
Next, it is also easy to show by using the last equation of
(\ref{eqn:diff-eq-RK}) that, if $y_4(x_1)\leq 0$ and
$y_3(x_1)<y_{min}$ for $0<{}^{\exists}x_1<1$, then $y_4(x)\leq 0$ for
$x_1\leq{}^{\forall}x\leq 1$, where $y_{min}$ is the global minimum of
$V(y_3)$ between $y_-$ and $y_+$. Thus, if $y_3$ starts from $y_+$ at
$x=0$ and reaches the region $y_3<y_{min}$ then $y_3$ cannot return to 
$y_+$ at $x=1$. Thirdly, combining this fact with (\ref{eqn:deV=0}),
we can show that there is no solution of the ($++$)-type which is
bounded in the region $y_-\leq y_3\leq y_+$. Actually,
(\ref{eqn:deV=0}) implies that if $y_3$ starts from $y_+$ at $x=0$ and
is bounded in the region $y_-\leq y_3\leq y_+$ then $y_3$ should reach
the region $y_3<y_{min}$, from which $y_3$ cannot return to $y_+$ at
$x=1$. Similarly, we can show that there is no solution of the
($--$)-type which is bounded in the region $y_-\leq y_3\leq
y_+$. Heuristically, the above mathematical statement can be easily
inferred if one interprets the last two equations of
(\ref{eqn:diff-eq-RK}) as the equation of motion for a particle
position $y_3(x)$ at time $x$. In this interpretation, the particle
receives the force due to the reversed potential $-(L/l)^2V(y_3)/2$
and the friction (or anti-friction) force $4y_2\dot{y}_3$. Moreover,
with this heuristic interpretation in mind, it seems likely that any
solution should be bounded in the region $y_-\leq y_3\leq y_+$ since
the reversed potential $-(L/l)^2V(y_3)/2$ is negative outside the
region $y_-\leq y_3<y_+$ and $V(y_3(0))=V(y_3(1))=0$. Finally,
combining this with the above mathematical statement, we can say at
least for moderate $\alpha$ that there are no solutions of ($++$)- and
($--$)-types.

Moreover, ($+-$)- and ($-+$)-types are equivalent to each other up
to the coordinate transformation $x\to 1-x$. Hence, ($-+$)-type 
solutions are all we have been seeking.

\begin{figure}
 \begin{center}
  \epsfile{file=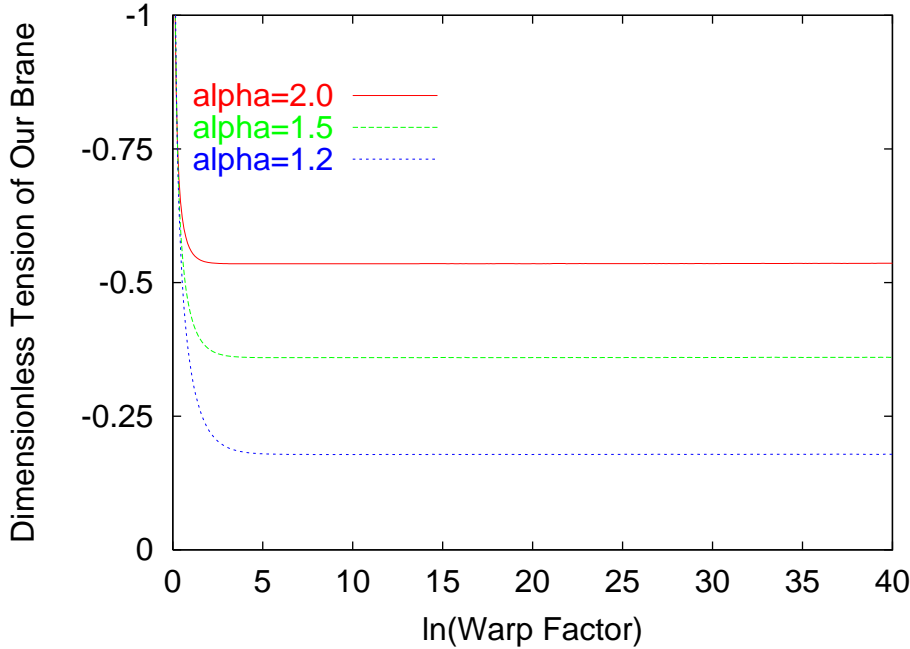,scale=1.0}
 \end{center}
\caption{
The relation between the warp factor $\phi$ given by
(\ref{eqn:warp-factor}) and the tension $\lambda$ of our brane at
$x=0$, which is given by (\ref{eqn:brane-tension}), for the
($-+$)-type solutions. The horizontal axis represents $\ln\phi$ and
the vertical axis represents $\lambda/(6\kappa^{-2}l^{-1})$. The
physical parameter is $\alpha=2.0$, $1.5$, $1.2$. 
As $\phi$ becomes large, $\lambda/(6\kappa^{-2}l^{-1})$ converges
quickly to the $\alpha$-dependent value given by
(\ref{eqn:tension-asymptotics1}). 
}
	\label{fig:tension1}
\end{figure}

\begin{figure}
 \begin{center}
  \epsfile{file=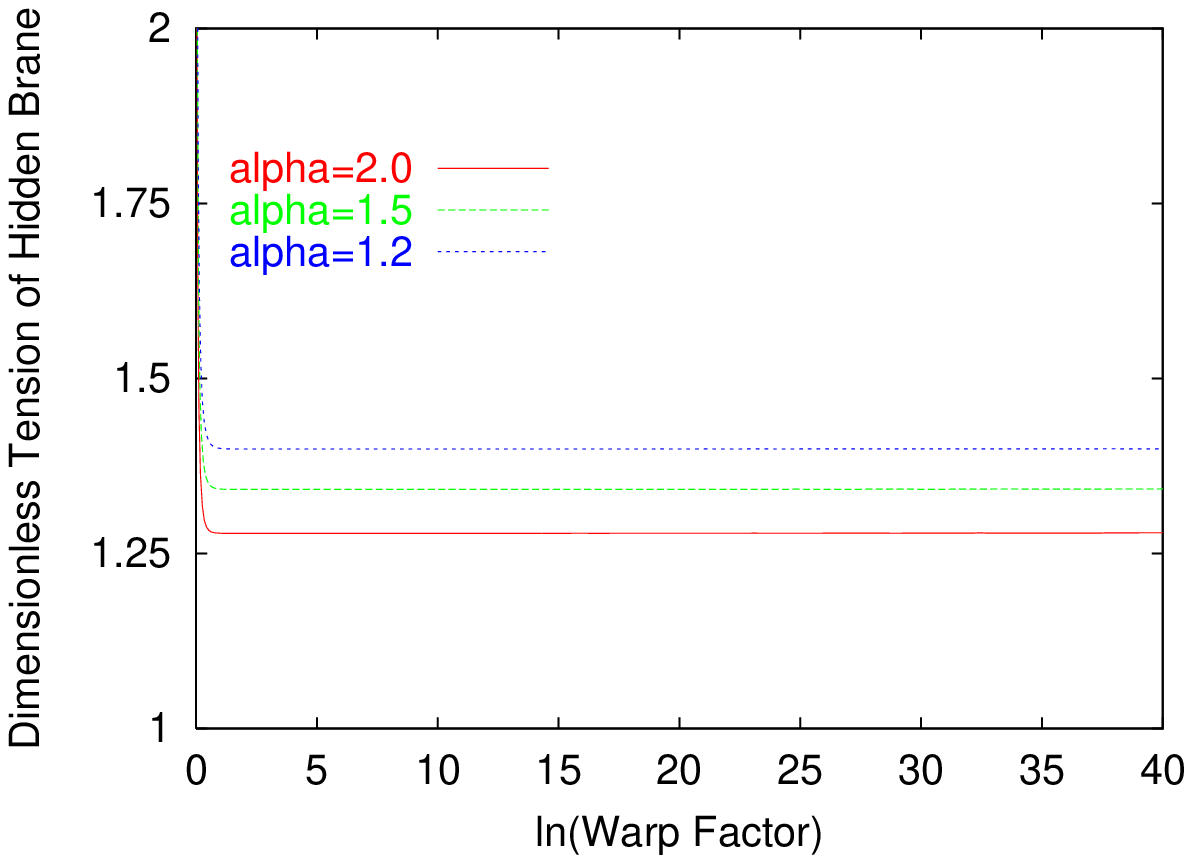,scale=1.0}
 \end{center}
\caption{
The relation between the warp factor $\phi$ given by
(\ref{eqn:warp-factor}) and the tension $\bar{\lambda}$ of the hidden
brane at $x=L$, which is given by (\ref{eqn:brane-tension}), for the 
($-+$)-type solutions. The horizontal axis represents $\ln\phi$ and
the vertical axis represents $\bar{\lambda}/(6\kappa^{-2}l^{-1})$. The 
physical parameter is $\alpha=2.0$, $1.5$, $1.2$. 
As $\phi$ becomes large, $\bar{\lambda}/(6\kappa^{-2}l^{-1})$
converges quickly to the $\alpha$-dependent value given by
(\ref{eqn:tension-asymptotics2}). 
}
	\label{fig:tension2}
\end{figure}

Figures~\ref{fig:tension1} and \ref{fig:tension2} show relations
between the warp factor $\phi$ given by (\ref{eqn:warp-factor}) and
the brane tension $\lambda$ or $\bar{\lambda}$ given by
(\ref{eqn:brane-tension}) for ($-+$)-type solutions. 
We can see that the tension of each brane converges quickly to an 
$\alpha$-dependent value when the warp factor becomes large. To be 
precise, 
%============< EQUATION >==============%
%
\begin{eqnarray}
 \lambda/(6\kappa^{-2}l^{-1}) & \to & \left\{
	\begin{array}{cc}
	-0.54 & (\alpha=2.0) \\
	-0.36 & (\alpha=1.5) \\
	-0.18 & (\alpha=1.2) 
	\end{array} \right.,	
	\label{eqn:tension-asymptotics1}\\
 \bar{\lambda}/(6\kappa^{-2}l^{-1}) & \to & \left\{
	\begin{array}{cc}
	1.28 & (\alpha=2.0) \\
	1.34 & (\alpha=1.5) \\
	1.40 & (\alpha=1.2)
	\end{array} \right.
	\label{eqn:tension-asymptotics2}
\end{eqnarray}
%======================================%
as $\phi\to\infty$. The convergence can be easily understood as
follows by the heuristic interpretation. We consider a particle in one
dimension whose position at time $x$ is $y_3(x)$ and which receives a
force due to the reversed potential $-(L/l)^2V(y_3)/2$ and the
friction (or anti-friction) force. The particle moves from $y_3=y_-$
to $y_3=y_+$ in the fixed duration $0\leq x\leq 1$. For a large $L/l$,
in order to satisfy this boundary condition, the particle should stay
near $y_3=y_{min}$ for a relatively long time since the reversed
potential is steep. In this case, the initial (or final) velocity
$y_4(0)$ (or $y_4(1)$, respectively) should be fine-tuned to a value
close to the 'escape velocity', which is roughly proportional to
$L/l$, against the reversed potential. Because of the relation
$y_2(0)+y_4(0)=0$ (or $y_2(1)+y_4(1)=0$, respectively) and the
junction condition (\ref{eqn:brane-tension}), the fine-tuning of
$y_4(0)$ (or $y_4(1)$, respectively) is equivalent to the fine-tuning
of $\lambda$ (or $\bar{\lambda}$, respectively). The required value of
$\lambda$ (or $\bar{\lambda}$, respectively) is almost independent of
$L/l$ since the required value of $y_4(0)$ (or $y_4(1)$, respectively)
is roughly proportional to $L/l$ as stated above. On the other hand,
when $y_3$ stays near $y_{min}$ with a very small velocity, $y_2$
should satisfy $y_2^2\approx -(L/l)^2V(y_{min})$ since the right hand
side of the last equation of (\ref{eqn:diff-eq}) should vanish
approximately. Hence, $y_1$ grows approximately linearly in $x$ with
the growth rate proportional to $L/l$ when $y_3$ stays near
$y_{min}$. Thus, the exponent of the warp factor should be roughly 
proportional to $L/l$. Actually, except for vicinities of the
boundaries, the numerical solutions are well approximated by
(\ref{eqn:approx-y1-y3}) and thus the warp factor is well approximated
by 
%============< EQUATION >==============%
%
\begin{equation}
 \phi \simeq \exp
	\left[\frac{L}{l}\sqrt{|V(y_{min})|}+y_- - y_+\right]. 
\end{equation}
%======================================%
Finally, combining the approximate linearity of the exponent of the
warp factor with the fine-tuning of the brane tension, we can conclude
that the brane tension should converge to a constant as the warp
factor becomes large. The limiting value of the brane tension should
depend on the parameter $\alpha$ since the 'escape velocity' depends
on $\alpha$. This conclusion is of course consistent with the
numerical result.

\begin{figure}
 \begin{center}
  \epsfile{file=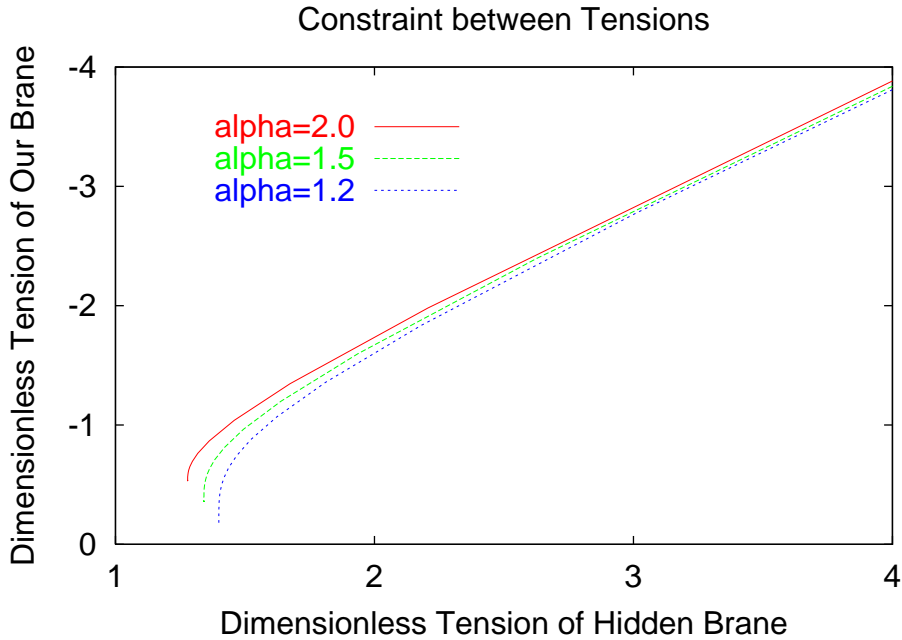,scale=1.0}
 \end{center}
\caption{
The relation between tension $\lambda$ of our brane at $x=0$ and the
tension $\bar{\lambda}$ of the hidden brane at $x=1$. The horizontal
axis represents $\bar{\lambda}/(6\kappa^{-2}l^{-1})$ and the vertical
axis represents $\lambda/(6\kappa^{-2}l^{-1})$. This relation can be
considered as a necessary condition for the system with two branes to
be static, or the condition for the four-dimensional cosmological
constant to vanish. The physical parameter is $\alpha=2.0$, $1.5$,
$1.2$. 
}
	\label{fig:tension3}
\end{figure}

Figure~\ref{fig:tension3} shows a relation between tension of our
brane at $x=0$ and that of the hidden brane at $x=1$. This relation
can be considered as a necessary condition for the system with two
branes to be static, or the condition for the four-dimensional
cosmological constant to vanish.

%============< FIGURE >==============%
%	plot-tension1.eps
%======================================%

%============< FIGURE >==============%
%	plot-tension2.eps
%======================================%

%============< FIGURE >==============%
%	plot-tension3.eps
%======================================%

Now let us give an argument that there is no static solution for 
$\alpha\leq 1$. The following argument is almost the same as that for
the non-existence of $(++)$- and $(--)$-type solutions. First, let us
consider the case $0<\alpha\leq 1$. In this case the potential
$V(y_3)$ has only one root $y=y_+$, and thus only $(++)$-type
solutions are allowed if any. However, since $V'(y_3)>0$ for $y_3\leq
y_+$, the condition (\ref{eqn:deV=0}) excludes $(++)$-type solutions
which are bounded in the region $y_3\leq y_+$. On the other hand,
since the reversed potential $-(L/l)^2V(y_3)/2$ is negative for
$y_3>y_+$ and $V(y_3(0))=V(y_3(1))=0$, the above heuristic
interpretation in terms of a particle motion strongly suggests that,
if there exists a solution of the differential equation, the solution
should be bounded in the region $y_3\leq y_+$. Hence, we can conclude
that there is no solution for $0<\alpha\leq 1$. Next, for 
$\alpha\leq 0$, there is no root of the potential $V(y_3)$. Hence, in
this case there is no way to satisfy the boundary condition
(\ref{eqn:y3=root}). Therefore, there is no solution for $\alpha\leq
1$, and it is actually enough to concentrate on the case $\alpha>1$ as
we did.

%======================================%
% Summary and discussions
%======================================%
\section{Summary and discussions}
	\label{sec:summary}

We have proposed a simple five-dimensional brane world model,
motivated by M-theory compactified on a six-dimensional manifold 
of small radius and an $S^1/Z_2$ of large radius. We have included
the leading-order higher curvature correction to the tree-level bulk 
action since in brane world scenarios the curvature scale in the bulk
may be comparable to the five-dimensional Planck scale and, thus, 
higher curvature corrections may become important. As a manageable
model of the bulk theory we have considered pure gravity including a 
$($Ricci-scalar$)^4$-correction to the Einstein-Hilbert action.

In this model theory, after a conformal transformation to the Einstein 
frame, we have numerically obtained static solutions, 
each of which consists of a positive tension brane and a negative
tension brane. The solutions are parameterized by a dimensionless
parameter $\alpha$ in the bulk theory and $L/l$, where $L$ is distance
between two branes and $l$ is a length scale determined by the
(negative) bulk cosmological constant. Several solutions are shown in
Figures~\ref{fig:y1-10}, \ref{fig:y1-20}, \ref{fig:y1-30},
\ref{fig:y3-10}, \ref{fig:y3-20} and \ref{fig:y3-30}.

The warp factor and tension of both branes have been calculated for
various values of $\alpha$ and $L/l$ and, by eliminating $L/l$, 
we have obtained two $\alpha$-dependent relations between the warp
factor and brane tensions. The existence of these relations implies
that, contrary to the original Randall-Sundrum model, the so called
radion is no longer a zero mode. In this sense, the present model is
similar to those in Refs.~\cite{GW,DFGK,gravity-stabilization}. The
two relations completely determine the brane tensions as functions of
the warp factor and are shown in Figures~\ref{fig:tension1} and 
\ref{fig:tension2}. From these figures we conclude that the tension of
our brane should be negative and that fine-tuning of the tension of
both branes is necessary for a large warp factor to explain the large
hierarchy between the Planck scale and the electroweak scale. To be
precise, the brane tensions should be fine-tuned with high accuracy to
values shown in equations (\ref{eqn:tension-asymptotics1}) and  
(\ref{eqn:tension-asymptotics2}).

Further, eliminating the warp factor from Figures~\ref{fig:tension1} 
and \ref{fig:tension2}, we have obtained a relation between the brane 
tensions. It is shown in Figure~\ref{fig:tension3} and can be
considered as a necessary condition for the system with two branes to
be static, or the condition for the four-dimensional cosmological
constant to vanish. Namely, unless this relation is satisfied, the
system cannot be static but becomes dynamical, regardless of initial
conditions (i.e. initial position of branes, initial velocity of
branes, and so on).

A stability analysis of solutions obtained in the present paper is an
important topic for future work. Here, we only offer a comment
concerning this subject: we cannot derive a correct effective action
by simply substituting the solutions into the action. Actually, if we 
substitute any static solutions into the action then the action
vanishes because of the Hamiltonian constraint. A simple illustration
of this fact is given in Appendix~\ref{app:action}.

Several extensions of the present model may also be of interest for
future work: (i) inflating brane solution; (ii) cosmological solution;
(iii) inclusion of Ricci tensor and Weyl tensor contributions to the
$R^4$-term in the action; (iv) inclusion of a $3$-form field and
modulus corresponding to the six-dimensional compactification. (i) It
is probably not difficult to extend the static solutions in the
present paper to inflating brane solutions. The relation between brane
tensions, corresponding to Figure~\ref{fig:tension3}, is expected to
become dependent on the four-dimensional cosmological constant induced
on branes as well as the model parameter $\alpha$. (ii) Extension to 
cosmological solutions should be possible. This should not be as
difficult as extension of the semiclassical solutions in
ref.~\cite{Mukohyama2000} to cosmological solutions. The latter seems
rather difficult because of the so called moving mirror
effect~\cite{BD}, which is non-local. On the other hand, the present
model has a locally defined Lagrangian density in five
dimensions. Hence, extension to the cosmological context is easier in
the present model than in the model of
ref.\cite{Mukohyama2000}. Moreover, the present model seems more 
realistic and, thus, worth while investigating in more detail. (iii) 
Although we have investigated effects of the
$($Ricci-scalar$)^4$-correction only, it would be desirable to
investigate effects of other forth-order curvature terms. Note,
however, that effects of forth-order Weyl terms are probably less
important insofar as we consider the metric (\ref{eqn:ansatz}) or
small perturbations around it, since the Weyl tensor vanishes for the
metric (\ref{eqn:ansatz}). (iv) In the present model we have
considered pure gravity in the bulk. However, more realistic model
should include a $3$-form field existing in the bosonic sector of
eleven dimensional supergravity as well as moduli fields due to the
compactification from eleven dimensions to five dimensions. In this
case, we may consider effects due to various eighth-order derivative
terms other than $R^4$ terms.

%%%%%%%%%%%%%%%%%%%%%%%%%%%%%%%%%%%%%%%%%%%%%%%%%%%%%%%%%%%%%%%%%%%%
%%%%%%%%%%%%%%%%%%%%%%%%%%%%%%%%%%%%%%%%%%%%%%%%%%%%%%%%%%%%%%%%%%%%
% Acknowledgements
%%%%%%%%%%%%%%%%%%%%%%%%%%%%%%%%%%%%%%%%%%%%%%%%%%%%%%%%%%%%%%%%%%%%
%%%%%%%%%%%%%%%%%%%%%%%%%%%%%%%%%%%%%%%%%%%%%%%%%%%%%%%%%%%%%%%%%%%%
\vspace{1cm}

The author would like to thank Professor W.~Israel for continuing
encouragement and careful reading of this manuscript. He would be
grateful to Professor J.~Yokoyama and Dr. T.~Shiromizu for helpful
discussions in the early stage of this work. In particular, the author 
would greatly appreciate Professor J.~Yokoyama for informing as to his 
interesting work on topological $R^4$ inflation~\cite{EKOY}. This work 
is supported by CITA National Fellowship and the NSERC operating
research grant.

%%%%%%%%%%%%%%%%%%%%%%%%%%%%%%%%%%%%%%%%%%%%%%%%%%%%%%%%%%%%%%%%%%%%
%%%%%%%%%%%%%%%%%%%%%%%%%%%%%%%%%%%%%%%%%%%%%%%%%%%%%%%%%%%%%%%%%%%%
% Appendix
%%%%%%%%%%%%%%%%%%%%%%%%%%%%%%%%%%%%%%%%%%%%%%%%%%%%%%%%%%%%%%%%%%%%
%%%%%%%%%%%%%%%%%%%%%%%%%%%%%%%%%%%%%%%%%%%%%%%%%%%%%%%%%%%%%%%%%%%%

\appendix

%======================================%
%<<<<< Estimate of the action >>>>>>>>>%
%======================================%

\section{Estimate of the action}
	\label{app:action}

By the ansatz (\ref{eqn:ansatz}), the action
(\ref{eqn:equivalent-action}) is reduced to 
%============< EQUATION >==============%
%
\begin{eqnarray}
 I & = & \int d^4x{\cal L},     \nonumber\\
 {\cal L} & = & \oint dw e^{-4A}\left\{\frac{2}{\kappa^2}
        \left[2\frac{d^2A}{dw^2}-5\left(\frac{dA}{dw}\right)^2\right] 
        - \left[\frac{1}{2}\left(\frac{d\psi}{dw}\right)^2
	+\tilde{U}(\psi)\right]\right\},
\end{eqnarray}
%======================================%
where the integration with respect to $w$ in this expression is over
the whole $S^1$, and 
%============< EQUATION >==============%
%
\begin{equation}
 \tilde{U}(\psi) = U(\psi) + f(\psi)
        \delta(w)+\bar{f}(\psi)\delta(w-L).
\end{equation}
%======================================%
From this reduced action, the following equations of motion are
derived. 
%============< EQUATION >==============%
%
\begin{eqnarray}
 3\left[2\left(\frac{dA}{dw}\right)^2-\frac{d^2A}{dw^2}\right] 
        + \kappa^2\left[\frac{1}{2}\left(\frac{d\psi}{dw}\right)^2 
        +\tilde{U}(\psi)\right]
        & = & 0,\nonumber\\
 e^{4A}\frac{d}{dw}\left(e^{-4A}\frac{d\psi}{dw}\right)
	-\tilde{U}'(\psi)=0.
        \label{eqn:reduced-eom}
\end{eqnarray}
%======================================%
These are equivalent to equations (\ref{eqn:E-eq}),
(\ref{eqn:matching-cond}) and (\ref{eqn:junction-cond}), provided that 
the identification $w\sim w+2L\sim -L$ is imposed. We can estimate the
value of the reduced action by using these equations. Actually, the
first of (\ref{eqn:reduced-eom}) reduces the effective Lagrangian
density ${\cal L}$ to 
%============< EQUATION >==============%
%
\begin{eqnarray}
 {\cal L} & = & \frac{1}{\kappa^2}
        \oint dw
        e^{-4A}\left\{2\left[2\frac{d^2A}{dw^2}
	-5\left(\frac{dA}{dw}\right)^2\right]
        +3\left[2\left(\frac{dA}{dw}\right)^2-\frac{d^2A}{dw^2}
	\right]\right\}\nonumber\\
 & = & \frac{1}{\kappa^2}\oint dw 
	\frac{d}{dw}\left(e^{-4A}\frac{dA}{dw}\right)
        = 0.
\end{eqnarray}
%======================================%
Therefore, the effective Lagrangian density vanishes if a 
solution of equations of motion is substituted. This fact can be
easily understood as follows~\footnote{
The author would thank Professor J.~Yokoyama for pointing out this.}:
in the static case without boundaries the Lagrangian of the system is
just minus the Hamiltonian, which should vanish because of the
Hamiltonian constraint.

The above arguments can be applied to the situation in ref.~\cite{GW}
by simply replacing $U(\psi)$, $f(\psi)$ and $\bar{f}(\psi)$ with
appropriate functions. Since the above analysis indicates a vanishing
effective potential, it is impossible to obtain a correct effective
potential for the so called radion, which corresponds to $L$ in the
above arguments, by simply substituting solutions to the action. It
seems that the non-vanishing effective potential obtained in
ref.~\cite{GW} merely measures an amount of inconsistency in their 
analysis. Nonetheless, their conjecture that the radion can be
stabilized by inclusion of a bulk scalar field seems
correct if the backreaction of the scalar field to the geometry is
sufficiently small~\cite{gravity-stabilization}.

%======================================%
%<<<<<<<<<<<< REFERENCES >>>>>>>>>>>>>>%
%======================================%

\end{document}